\documentclass[%
 aapm,
 mph,%
 amsmath,amssymb,
 reprint,%
]{revtex4-2}

\usepackage{graphicx}
\usepackage{dcolumn}
\usepackage{bm}
\usepackage{comment}
\usepackage[capitalize]{cleveref}

\usepackage{physics}
\usepackage{xcolor}

\begin{document}


\title[Implementation security in quantum key distribution]{Implementation security in quantum key distribution}
	\author{Víctor Zapatero$^{1,2,3}$}
    \author{Álvaro Navarrete$^{1,2,3}$}
	\author{Marcos Curty$^{1,2,3}$}
 
\affiliation{$^{1}$Vigo Quantum Communication Center, University of Vigo, Vigo E-36310, Spain}

\affiliation{$^2$Escuela de Ingeniería de Telecomunicación, Department of Signal Theory and Communications, University of Vigo, Vigo E-36310, Spain}

\affiliation{$^3$AtlanTTic Research Center, University of Vigo, Vigo E-36310, Spain}

\begin{abstract}
The problem of implementation security in quantum key distribution (QKD) refers to the difficulty of meeting the requirements of mathematical security proofs in real-life QKD systems. Here, we provide a succint review on this topic, focusing on discrete variable QKD setups. Particularly, we discuss some of their main vulnerabilities and comment on possible approaches to overcome them.

\end{abstract}

\maketitle

\section{Introduction}
Almost forty years after its conception~\cite{bennett1984proceedings}, quantum key distribution (QKD) has become a mature technology excelling among quantum information applications, with dedicated companies providing QKD services~\cite{IdQuantique,Toshiba,qubridge,ThinkQuantum}, metropolitan QKD networks being deployed around the globe~\cite{peev2009secoqc,xu2009field,sasaki2011field,stucki2011long,dynes2019cambridge,yang2021all}, record secret key rates and transmission distances being achieved over optical fibres~\cite{grunenfelder2023fast,li2023high,boaron2018secure,chen2021twin,wang2022twin}, and a space-ground integrated QKD backbone being extended over thousands of kilometres~\cite{chen2021integrated} in China.

The crucial cryptographic advantage of QKD is that it allows for information-theoretically secure key exchange through an insecure channel~\cite{portmann2022security}, which provides a unique solution for long-term communication security. More specifically, while the security of standard public-key cryptosystems presumes computational limitations on a potential adversary, mathematical QKD security proofs rely on fundamental properties of quantum mechanics ---such as the unclonability~\cite{wootters1982single} of quantum states or quantum entanglement~\cite{nielsen2010quantum}--- together with models of the behaviour of the physical devices involved in a QKD protocol~\cite{curty2021quantum,xu2020secure}. To put it another way, QKD security stems from unique physical-layer properties, which is a double-edged sword: on the one-hand, it eliminates the need for computational assumptions that become increasingly weaker over time, but on the other hand, it is significantly dependent on the proper characterization of the quantum communication hardware. The latter observation gives rise to a variety of concerns regarding the security of QKD implementations, and in this review we unravel some of the intricacies of this topic, specifically for the case of discrete variable QKD.

As is customary in the literature on QKD, in what follows we often refer to the honest parties trying to establish a secret key as Alice and Bob, and to the eavesdropper in the quantum channel as Eve.
\section{Implementation Security}
To state it shortly, if the assumptions invoked by the security proof of a QKD protocol are not met exactly in a real implementation, the security of the implementation does not follow. In particular, QKD security proofs typically rely on mathematical models of the QKD equipment, which opens the door for mis-characterization loopholes. This is problematic given the complexity of QKD hardware and our engineering limitations, which prevent us from fully controlling the operation of the devices. On top of it, faithfully describing the functioning of a QKD transmitter or receiver with which an adversary can arbitrarily interact is a major challenge, as illustrated by the rich literature on quantum hacking~\cite{xu2020secure,jain2016attacks} and the variety of quantum hacking attacks successfully implemented~\cite{zhao2008quantum,lydersen2010hacking,gerhardt2011full,weier2011quantum,jain2011device,pang2020hacking}. For these reasons, rigorously quantifying the level of security of QKD implementations is a difficult task, an issue we refer to as the \textit{implementation security} of QKD. In fact, the problem of implementation security is also a threat for conventional cryptosystems, but it is probably more severe for QKD, being this an eminently hardware solution.

Fortunately, outstanding theoretical and experimental efforts are being made to address this problem. For at least two decades, theorists have been
devising new QKD protocols and security proofs more and more robust to experimental imperfections, and experimentalists have developed special-purpose hardware for optical quantum communications, getting closer and closer to the specifications of the theoretical security proofs.

In a deeper level, the security proofs follow the so-called real-world vs ideal-world paradigm~\cite{portmann2022security}, in which the security of a protocol is quantified by measuring its distinguishability from an ideal execution of the cryptographic task under consideration. Particularly, the ideal functionality of QKD is defined by the following two requirements on the output key: it must be uniformly distributed and uncorrelated from the knowledge of Eve (secrecy requirement), and Alice and Bob must hold identical copies of it (correctness requirement). In order to evaluate the distinguishability of a QKD protocol from this ideal functionality, very diverse assumptions are typically presumed. From the implementation point of view, an inevitable premise seems to be the proper isolation of the parties' QKD labs. Indeed, since the exchange of quantum signals requires that the labs be connected to the outside world through a quantum channel, this assumption already gives rise to implementation issues, with information being potentially leaked in different ways, both active and passive. Furthermore, information leakage (IL) can also occur through classical channels, via \textit{e.g.} electromagnetic radiation, power consumption, or covert channels in sabotaged QKD hardware inter alia.

Generally speaking, passive information leakage in the quantum channel arises due to the mis-modelling of either the quantum states prepared in the QKD transmitter or of the quantum measurements performed in the QKD receiver. A paradigmatic example is given by the so-called photon-number splitting attack against the ideal BB84 protocol, originally suggested in~\cite{huttner1995quantum} and extensively discussed in \textit{e.g.}~\cite{brassard2000limitations}. Essentially, the attack is based on photon counting and may reveal Eve the bit values redundantly encoded in multi-photon pulses without introducing errors. This is so because multi-photon pulses provide Eve with perfect copies of the state of the single-photon pulses, thus bypassing the limitations imposed by the no-cloning theorem~\cite{wootters1982single}. Security-wise, the problem is not the presence of multi-photons in the states emitted by the laser, but rather the assumption that individual photons are being prepared. Indeed, a more realistic approach accounting for the presence of multi-photons consists of modelling the laser pulses as phase-randomized weak coherent states~\cite{inamori2007unconditional} when the laser is driven under gain-switching conditions. Within this model, the threat of photon-number splitting attacks can be overcome by using the decoy-state method~\cite{hwang2003quantum,lo2005decoy,wang2005beating}, which consists on randomly alternating among various intensities in the QKD transmission. Irrespectively of Eve's attack, the detection statistics of the different intensities allow to tightly estimate the detection statistics of the single-photon pulses, from which the provably secure key length can be calculated.

In the following, we address several other implementation security breaches, separately discussing receiver vulnerabilities and transmitter vulnerabilities in different sections. Coming next, we shortly present the device-independent (DI) approach to QKD and the threat of malicious/sabotaged devices, to finally end up the review with some prospects and concluding remarks.

\section{Receiver vulnerabilities}

The receiver is widely regarded as the most vulnerable element of a QKD scheme. This is mainly attributable to its need to be open to interactions for signal measurement purposes, thereby providing Eve opportunity to readily interact with it and manipulate its operation. Moreover, inevitable imperfections offer an exploitable scenario for Eve to enhance her malicious interventions. 

For instance, most security analyses assume that all detectors at the receiver have the same detection efficiency. This facilitates the merging of detection inefficiency and channel transmission loss under adversary control, allowing the use of conventional security-proof techniques in a simpler model with higher effective channel loss and ideal single-photon detectors (SPDs). In general, however, the detector's response to a photon often depends on various degrees of freedom ---\textit{e.g.} the spatial mode, frequency, or arrival time--- which may not intentionally play a role in the encoding. This allows Eve to perform \textit{e.g.} a fake-state attack~\cite{makarov2005faked,makarov2006effects}, a particular type of intercept-and-resend attack in which Eve randomly measures Alice's signals in one of the BB84 bases ---if we consider this particular protocol--- and then resends the state corresponding to the inverse of her bit guess in the opposite basis at a time when Bob's detector associated with that inverse bit has a significantly lower efficiency than the other detector. If Eve's and Bob's measurement bases are not in agreement, the signal resent by Eve will only collide with the low-efficiency detector and rarely trigger it. If their bases match, the signal will be distributed between the two detectors, but the one associated with the opposite of Eve's bit guess will be activated with lower probability. Eve could also perform a time-shift attack~\cite{qi2007time,zhao2008quantum} by simply shifting the arrival time at Bob's side of each signal sent by Alice to a time instance where the sensitivity of one detector far exceeds that of the other. In the most extreme scenarios ---where one detector is completely insensitive while the other remains receptive--- either of these attacks would allow Eve to learn the entire secret key without being detected.

SPDs might also suffer from backflash radiation, whereby photons are emitted upon a detection event due to the recombination of electrons with holes within the avalanche breakdown region~\cite{kurtsiefer2001breakdown,pinheiro2018eavesdropping}. While the backflash photons carry no information about the detected states, they have the potential to reveal information about the internal state of the receiver's components that they encounter while travelling to the channel, which may lead to IL about the final key. Indeed, IL about the receiver's internal configuration can even be actively induced by Eve via a Trojan-horse attack~\cite{sajeed2017invisible,jain2014risk} (THA). In a THA, Eve injects bright light into the receiver ---usually opting for a wavelength that avoids detector triggering at Bob--- which then undergoes reflections within the internal components of Bob's receiver, thereby conveying sensitive information to the channel.

The THA is an example of an attack in which Eve not only exploits existing imperfections to enhance her eavesdropping capabilities, but she interacts with Bob's devices to create new vulnerabilities. The best known instance of this kind is the so-called blinding attack~\cite{makarov2009controlling,gerhardt2011full,lydersen2010hacking,lydersen2011controlling,huang2016testing}, in which Eve injects strong light into the receiver's single-photon avalanche diodes (SPADs) to make them operate in linear mode rather than Geiger mode. As a result, the detectors no longer function as SPDs but as conventional photodiodes. This can be later used to manipulate the detectors' clicking behaviour by sending them classical light of properly chosen intensity.

Indeed, the blinding attack can be regarded as an instance of a broader class of intercept-and-resend attacks called detector-control attacks, where Eve does not try to replicate Alice's original states, but instead resends light pulses ---classical or quantum--- that manipulate Bob's detections such that they can only produce a click when the signals resent by Eve match those of Alice. Another example is the after-gate attack~\cite{wiechers2011after}, which targets gated SPADs. Here, Eve forces Bob's detectors to operate in linear mode by sending him strong signals just after the gating process, when the detectors are no longer in Geiger mode.

The \textit{dead time} of the SPDs may also open up potential vulnerabilities. Typically, the receiver only acknowledges detections that occur within a certain time window to minimise dark counts. This allows Eve to send light pulses to Bob before Alice's transmitted pulses reach the receiver, with the aim of triggering certain detectors at Bob's side~\cite{weier2011quantum} before the window is open. Since these detectors need time to reset, they cannot be triggered again in the same transmission round by Alice's original signals. Therefore, Eve can be confident that any subsequent detection announcement in that round must be attributed to the remaining operative detector(s).

Lastly, it has been demonstrated~\cite{bugge2014laser,makarov2016creation} that Eve could even use laser illumination to irreversibly modify some characteristics of the detectors ---such as their detection efficiency and dark count rate--- thereby creating permanent exploitable vulnerabilities in the receiver.

The previous examples represent common vulnerabilities in QKD receivers that require comprehensive risk mitigation strategies. Hardware-wise, a rigorous device calibration of the receiver's components is essential. For this, routine calibration intervals ---monitoring parameters such as current, voltage, temperature, and detection efficiency--- of Bob's detectors might help to ensure their correct functioning throughout the QKD session. To defend against invasive attacks from Eve ---such as blinding attacks or THAs--- optical isolators, circulators, and narrow spectral filters are often used to restrict any incoming light to the appropriate wavelength in Bob's receiver. In addition, optical fuses and power limiters can be used to limit the intensity of the incoming light from the channel. Moreover, the use of watchdog detectors with different response times also enables continuous monitoring of potential light-injection attempts by Eve.

An additional countermeasure against active IL is the implementation of passive receivers~\cite{boaron2018secure,rusca2018security}. This type of receiver passively distributes the incoming signals across the measurement bases, preventing the leakage of basis information to Eve via a THA.

With respect to the detector efficiency mismatch, a practical hardware solution is to use a four-state QKD receiver~\cite{nielsen2001experimental,lagasse2005secure} equipped with a single detector. In such a scheme, Bob randomly chooses not only the measurement basis, but also the bit to be measured. Indeed, this four-state idea can also be extended to two-detector receivers~\cite{qi2007time}, being in this case similar to randomly switching between the two detectors in every round. This guarantees that the detection efficiency associated to each bit value is the same.

From a theoretical point of view, considerable effort has been devoted to characterise the vulnerabilities of the devices and to incorporate these imperfections into the security proofs, a prime example being the GLLP framework~\cite{gottesman2004security}. Specifically, a preliminary security analysis against backflash radiation has been discussed \textit{e.g.} in~\cite{pinheiro2018eavesdropping}. Also, as demonstrated in~\cite{fung2010security,lydersen2008security,winick2018reliable,bochkov2019security,ma2019operational}, a well-characterized, stable detection-efficiency mismatch can be incorporated into the security proof, albeit under the premise that Eve cannot add photons to the transmitted signals. Remarkably, recent numerical~\cite{zhang2021security} and analytical~\cite{trushechkin2022security} security analyses have effectively removed this assumption.

Withal, there exists an ultimate solution that rules out all security loopholes concerning the receiver units, namely, interference-based QKD, such as measurement-device-independent~\cite{lo2012measurement} (MDI) or twin-field~\cite{lucamarini2018overcoming} (TF) QKD.

\subsection{Interference-based QKD}
\begin{figure}
    \centering
    \includegraphics[width=0.65\columnwidth]{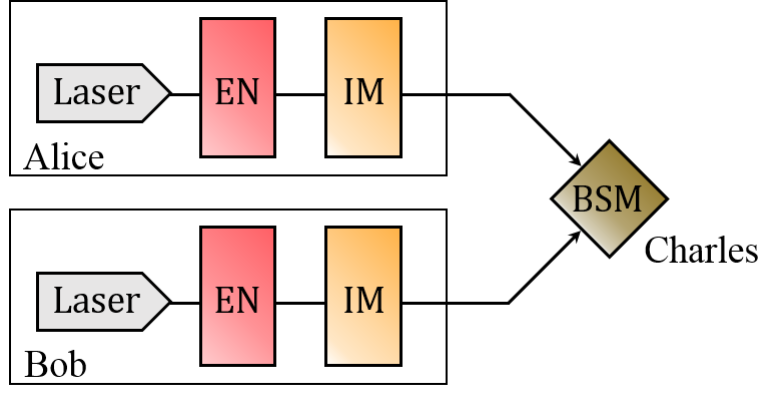}
    \caption{Illustration of an MDI-QKD setup. Alice and Bob send decoy-state BB84 pulses to an untrusted intermediate party, who performs a Bell-state measurement (BSM). EN: bit/basis encoder; IM: intensity modulator; BSM: Bell-state measurement.\label{fig:MDI}}
\end{figure}

In MDI-QKD both Alice and Bob play the role of transmitters, sending BB84 states to an intermediate party ---say, Charles--- who jointly measures the incoming signals and publicly announces the measurement results (see~\cref{fig:MDI}). The key feature of MDI-QKD is that its security does not rely on the integrity of Charles, who may even be Eve. Essentially, this is because MDI-QKD can be seen as a time-reversed Einstein-Podolsky-Rosen (EPR)-based QKD protocol~\cite{bennett1992quantum,biham1996quantum}. 

In an EPR-based protocol, Alice and Bob each prepares an EPR pair and send one half to Charles, who performs a Bell state measurement for entanglement swapping. Subsequently, Charles publicly announces the measurement results, and Alice and Bob randomly perform $Z$ and $X$ measurements ---defined by the qubit states $\{\ket{0},\ket{1}\}$ and $\{(\ket{0}+\ket{1})/\sqrt{2}, (\ket{0}-\ket{1})/\sqrt{2}\}$, respectively--- on their respective halves. By analysing the measurement statistics, Alice and Bob can detect any dishonest behaviour from Charles. Notably, this feature remains unchanged if Alice and Bob proceed with their local measurements before sending their respective EPR halves to Charles ---effectively preparing these qubits in BB84 states--- which is the basis of MDI-QKD.

In practice, Charles' measurement setup can be implemented with basic linear optical elements ---such as beamsplitters and polarising beamsplitters--- and threshold detectors, while the transmitters' are identical to those used in standard decoy-state QKD implementations. Due to its immunity to detector side channels, together with its practicality and good performance, MDI-QKD has given rise to numerous theoretical studies. These include refined parameter estimation methods~\cite{zhou2016making}, finite-key security analyses~\cite{ma2012statistical,curty2014finite} and alternative schemes using different encoding approaches~\cite{tamaki2012phase,ma2012alternative}. In fact, several experimental implementations have been reported in the last decade~\cite{rubenok2013real,tang2014experimental,wang2015phase,yin2016measurement,comandar2016quantum,wei2020high,liu2023experimental}.

A noteworthy variant of interference-based QKD is twin-field (TF) QKD~\cite{lucamarini2018overcoming}. It is well known that point-to-point QKD configurations have inherent limitations, causing the secret key rate to scale at best linearly with the transmittance of the quantum channel between the two parties. Unfortunately, despite its intermediate-node configuration, MDI-QKD is unable to overcome this limit due to the requirement of two-photon interference at the central node. In order to surpass this limitation, various theoretical solutions have been proposed, including for instance quantum repeaters~\cite{briegel1998quantum,duan2001long,sangouard2011quantum} or MDI-QKD protocols incorporating quantum memories~\cite{abruzzo2014measurement} or quantum non-demolition measurements~\cite{azuma2015all}. However, these solutions are impractical with existing technology. 

In this light, TF-QKD emerges as a practical solution that overcomes linear scaling through a setup that is implementable with current technology. Sharing its mid-node structure and immunity to detector side channels with MDI-QKD, the key rate of TF-QKD scales with the square root of the channel transmittance, thus essentially doubling the maximum achievable distance when compared to MDI-QKD. This feature can be attributed to the entanglement swapping operation at the central node, which is based on single-photon interference. In other words, the arrival of a single photon at the central node (be it from Alice or from Bob) is enough to distill a key. This advantage, nonetheless, comes at the cost of requiring strict phase stability.

In the last years, numerous TF-QKD variants have refined the original idea to enhance its practicality and performance~\cite{curty2019simple,ma2018phase,lin2018simple,wang2018twin}, many of them having already witnessed experimental implementations~\cite{minder2019experimental,wang2019beating,liu2019experimental,zhong2019proof,fang2020implementation,chen2020sending,liu2021field,chen2021twin,chen2022quantum,pittaluga2021600,wang2022twin,liu2023experimental}. Interestingly, a recent solution has mitigated the need for phase stability~\cite{zeng2022mode,zhu2023experimental}, thus greatly simplifying its practical deployment. Overall, this class of QKD protocols is becoming the standard for long-distance fibre communication. Indeed, this assertion is supported by the current record transmission distance over fibre~\cite{liu2023experimental}, which potentially delivers a positive asymptotic key rate for a link of 1002 km.

\section{Transmitter vulnerabilities}
Real-world transmitters also suffer from inevitable imperfections that are frequently ignored in security proofs. Among them, the most obvious may be encoding errors or state preparation flaws (SPFs). These refer to subtle deviations between the actually transmitted states and the idealized states considered in the theoretical models, and may arise due to the faulty construction of the encoding devices or finite precision limitations. In certain protocols, such as the BB84 scheme, SPFs often enable Eve to partially discern between the two bases. While the GLLP analysis~\cite{gottesman2004security} addresses SPFs, its performance deteriorates significantly at relatively short distances even for minor imperfections. Alternatively, the so-called loss-tolerant (LT) protocol~\cite{tamaki2014loss,mizutani2015finite,curras2021finite} resolves this problem by rendering the protocol almost insensitive to SPFs. For this, SPFs must be well characterized, and they must not take the transmitted single-photons out of their original qubit space.

Another instance of faulty state preparation can be found in decoy-state QKD protocols. The effective implementation of the decoy-state method requires precise knowledge and temporal stability of the intensities of Alice's transmitted pulses. However, practical transmitters often encounter minor intensity fluctuations that must be incorporated into the security analysis~\cite{mizutani2015finite}. In addition, it is crucial for the method's integrity that the optical phase of each coherent pulse appears uniformly random to any potential eavesdropper. A conventional approach for achieving this involves gain-switching the laser source~\cite{boaron2018secure,yuan2007unconditionally,dixon2008gigahertz,liu2010decoy,lucamarini2013efficient,valivarthi2017cost,yuan201810}. In doing so, each pulse is triggered by a spontaneous emission event, leading to an optical phase that is essentially random. Unfortunately, inherent imperfections in the devices can hinder the achievement of a perfectly uniform phase distribution~\cite{xu2012ultrafast,abellan2014ultra}. To solve this, recent techniques~\cite{nahar2023imperfect,curras2022security} permit to accommodate this kind of phase imperfections into the security proofs. Another frequently adopted solution involves utilising an external phase modulator~\cite{zhao2007experimental,sun2012experimental,sibson2017chip,bunandar2018metropolitan}. However, this requires a fresh supply of random numbers to feed the modulator. Worse still, only a discrete set of optical phases can be selected, thereby violating the fundamental assumption of the decoy-state method. To address this challenge, more sophisticated decoy-state analyses~\cite{cao2015discrete,curras2021twin} have been proposed to restore the security even when utilising a finite number of phases. Furthermore, these ideas have been extended to encompass the general scenario where the phase of each pulse follows an arbitrary ---continuous or discrete--- probability density function~\cite{sixto2023secret}.

As for the receivers, imperfections in the transmitters can also be actively induced by eavesdropping attacks. For example, in a laser seeding attack Eve deliberately manipulates the intensity~\cite{huang2019laser} and/or the phase~\cite{sun2015effect} of Alice's transmitted signals by introducing light into her laser source, thereby triggering stimulated emission. In a similar way, she could inject light to irreversibly damage the transmitter~\cite{huang2020laser}, permanently increasing the intensity of Alice's emitted signals.

\begin{figure}
    \centering
    \includegraphics[width=0.75\columnwidth]{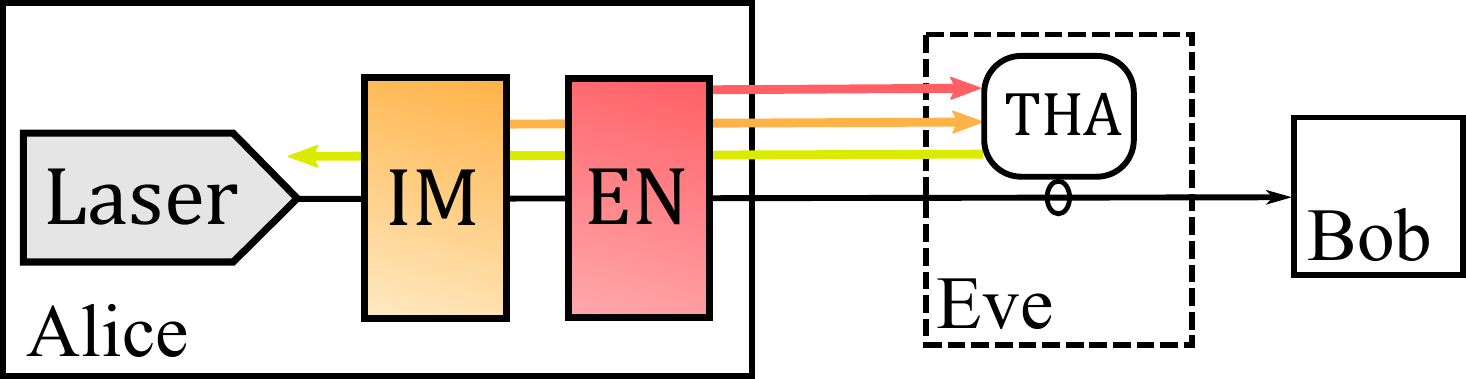}
    \caption{Schematic diagram of a THA. Eve injects a photonic state into Alice's transmitter and analyses the back-reflected light, which can provide information about the internal configuration of Alice's bit/basis encoder (EN) and intensity modulator (IM).\label{fig:THA}}
\end{figure}

Also crucial for the transmitter's integrity is to prevent any kind of IL. Passive generation of IL may be produced \textit{e.g.} via mode dependencies in Alice's transmitter~\cite{nauerth2009information,duplinskiy2021bounding,bourassa2022measurement}. This occurs when Alice's settings become encoded in other degrees of freedom associated with the emitted light ---such as frequency and time--- or through unintended emissions of a different nature ---such as unwanted electromagnetic radiation, power consumption, or even apparatus-generated noise--- resulting in the transmission of higher-dimensional quantum states. From this perspective, IL can be regarded as a specific manifestation of imperfect state preparation, in which the dimensionality of the transmitted states increases, thereby creating a side channel that enhances the distinguishability of the signals and thus the effectiveness of Eve's attacks. For example, Alice's signals might become linearly independent, in so allowing Eve to perform an unambiguous state discrimination attack. Needless to say, this would severely restrict the performance of the protocol. Moreover, Eve could actively provoke a similar effect \textit{e.g.} by launching a THA~\cite{gisin2006trojan,vakhitov2001large} against the transmitter (see~\cref{fig:THA}), or by establishing a dependency between the encoding of Alice's signals and their frequency~\cite{pang2020hacking}. In fact, this could be executed in tandem with any of the aforementioned laser injection attacks.

IL may also manifest through pulse correlations. To enhance the secret key rate, modern QKD systems aim to increase the repetition rate of the sources, achieving up to gigahertz repetition rates in current experimental implementations~\cite{schmitt2007experimental,frohlich2017long,yuan201810,boaron2018secure}. However, operating at high speed typically introduces memory effects in the modulators and electronics that control them~\cite{grunenfelder2020performance,kobayashi2014evaluation,roberts2018patterning,yoshino2018quantum}, resulting in correlations between optical pulses. This means that the state of a quantum signal emitted by the source at a given instant may depend not only on the settings chosen by Alice at that round, but also on previous setting choices. As a result, the settings used to encode each quantum signal can be partially inferred from the quantum states of subsequent signals, acting as a side channel. 

Securing the source against IL is challenging. Nonetheless, significant effort has been devoted by theorists in developing security analyses that encompass their various manifestations. For example, recent security analyses~\cite{pereira2019quantum,navarrete2021practical,pereira2023modified} have successfully accommodated multiple source imperfections ---as mode dependencies, SPFs, and pulse correlations--- using the novel reference technique (RT) framework~\cite{pereira2020quantum}. Essentially, the method considers some reference states that closely resemble the actual states transmitted in the protocol, but whose simpler structure facilitates the estimation of Eve's side information. Then, by exploiting the similarity between these two sets of states ---\textit{e.g.} via a Cauchy-Schwarz type inequality~\cite{pereira2020quantum}--- one can estimate Eve's side information from the detection statistics of the actual transmitted states.

\begin{figure}
    \centering
    \includegraphics[width=\columnwidth]{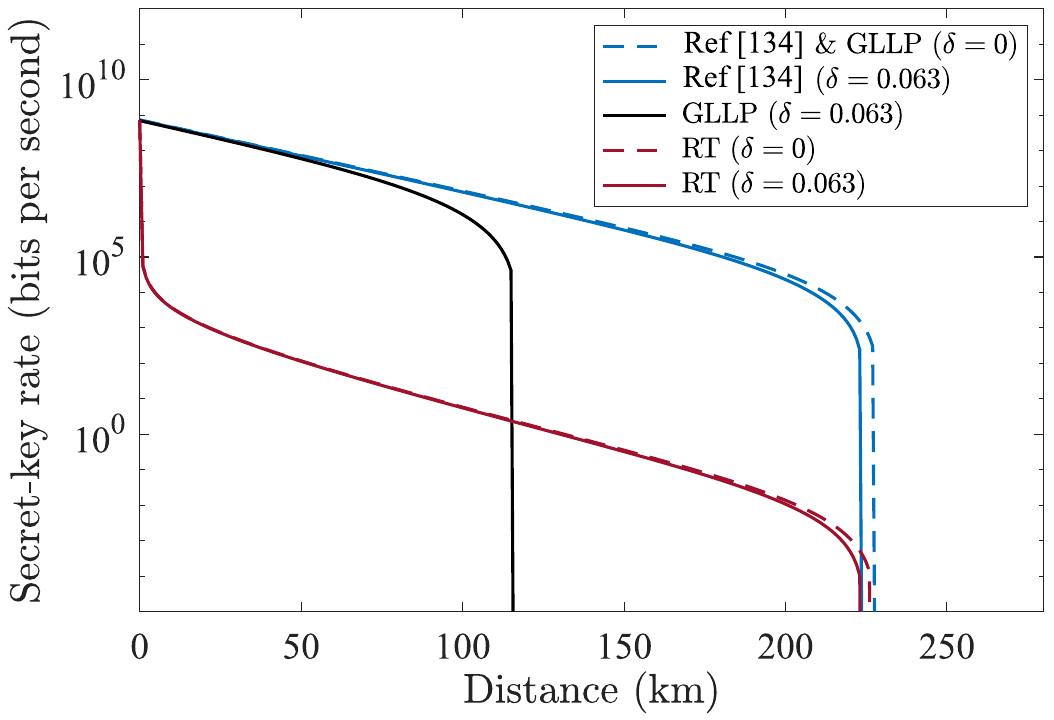}
    \caption{Rate-distance performance of the BB84 protocol with the recent security proof in~\cite{curras2023security}, compared with that of the original quantum coin (GLLP) \cite{gottesman2004security} and reference technique (RT) \cite{pereira2020quantum} analyses. The figure considers a lower bound on the fidelity between the actual leaky states and the reference states (without side channel) of $\epsilon=10^{-6}$. The parameter $\delta$ represents the magnitude of the qubit flaws, and an attenuation coefficient $\alpha=0.2$ dB/km is assumed. The figure has been adapted from~\cite{curras2023security}.\label{fig:CoinProtocol}}
\end{figure}

The high flexibility and tolerance to source imperfections of the RT comes with certain trade-offs. Specifically, it imposes a sequential execution of the protocol where Alice emits a pulse only after Bob has measured the preceding one. It is worth mentioning though that simple relativistic arguments can be invoked to relax this constraint~\cite{pereira2023modified,curras2023security}, and parallel executions of the protocol can mitigate its effects. Fortunately, a recent solution proposed in~\cite{curras2023security} has effectively overcome this constraint by reintroducing the quantum coin idea~\cite{gottesman2004security,lo2006security} and integrating it with the rejected data analysis of the LT protocol~\cite{tamaki2014loss} alongside random sampling techniques~\cite{curras2021finite}. By doing so, the method in~\cite{curras2023security} effectively considers the inherent distinction between qubit SPFs and side channels ---unlike the original GLLP analysis--- resulting in enhanced robustness to channel loss (see ~\cref{fig:CoinProtocol}). Withal, a major assumption underlying all of the techniques discussed in this section is that a fidelity-type bound between the real leaky systems and their idealized counterparts is satisfied, its precise experimental verification still being an open problem.

As mentioned above, pulse correlations can be regarded as a particular manifestation of IL and, therefore, some of the aforementioned techniques already allow to accommodate them into the security analysis. Setting-choice-independent correlations ---that is, pulse correlations unaffected by the setting choices made during state preparation--- have been extensively explored in the literature~\cite{nagamatsu2016security,mizutani2019quantum}. Likewise, recent works have addressed correlations in the bit/basis encoding for the more comprehensive setting-choice-dependent scenario~\cite{pereira2020quantum,pereira2023modified}, as well as intensity correlations in decoy-state QKD protocols~\cite{yoshino2018quantum,zapatero2021security,sixto2022security}. Moreover, correlations between the optical phases of coherent pulses emitted by a gain-switched laser have been addressed~\cite{curras2022security} recently.

Security proofs that contemplate THAs against the transmitter include~\cite{lucamarini2015practical,tamaki2016decoy,wang2018finite,pereira2019quantum,navarrete2021practical,tan2021chip,navarrete2022improved,ding2023improved,pereira2023modified}. Naturally, the transmitter may be experimentally safeguarded from THAs by installing optical isolators, narrow spectral filters, optical fuses, and power limiters at its output~\cite{todoroki2004optical,zhang2021securing,ponosova2022protecting}. Moreover, monitor detectors could be employed to measure the intensity of the incident light. However, these hardware solutions are limited, and achieving a perfect level of isolation is impossible. Hence, it is necessary to combine the above techniques (capable of upper-bounding the maximum intensity of Eve's back-reflected light) with appropriate security proofs that accommodate IL.

\begin{figure}[!htbp]
    \centering
    \includegraphics[width=\columnwidth]{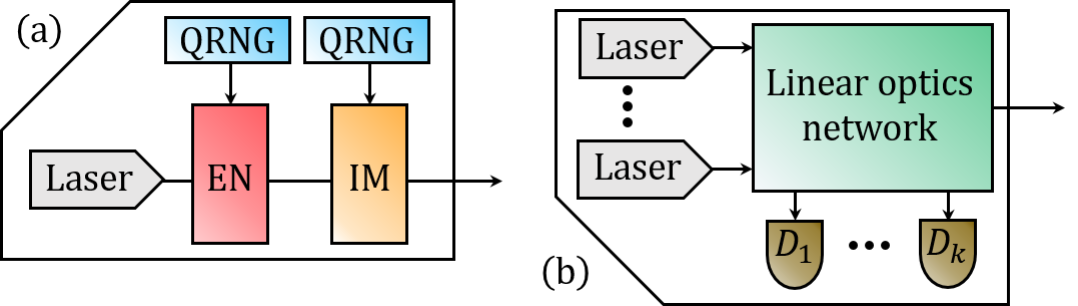}
    \caption{Comparison of (a) active and (b) passive decoy-state QKD transmission. In the active setup, an encoding device (EN) and an intensity modulator (IM) are externally driven by quantum random number generators (QRNGs) to implement the protocol settings. In the passive setup, one or more laser sources generate pulses that interfere in a linear optics network. Based on the detection outcomes observed in the photo-detectors ($\{D_{j}\}$), different signal states can be post-selected.\label{fig:passive}}
\end{figure}

Importantly as well, since, as stated above, memory effects in the modulators pose an IL threat, a modulator-free approach to QKD has also been extensively explored, known as passive QKD (see Fig.~\ref{fig:passive}). In this regard, the works in~\cite{wang2023fully,zapatero2023fully} proposed a QKD transmitter passively implementing both the encoding of BB84 states and the preparation of decoy intensities using linear optical elements and coherent light, combining earlier ideas presented in~\cite{curty2010passive,curty2010passive_2}. In fact, these papers were soon followed by proof-of-principle experimental demonstrations~\cite{lu2023experimental,PhysRevLett.131.110801} and novel proposals for fully passive TF-QKD~\cite{wang2023fully_TF} and MDI-QKD~\cite{li2023passive}. It is nevertheless worth mentioning that, although the passive approach may remove all modulation side-channels, passive QKD transmission relies on the post-selection of the desired protocol states through one or more (typically classical) measurements, possibly opening other security loopholes that have not been studied in detail yet~\cite{zapatero2023fully,zapatero2023finite}. As an example, a passive source might be vulnerable to intrusions similar to laser-seeding~\cite{huang2019laser} or laser-damage~\cite{huang2020laser} attacks, but with light injections potentially targeted at both the laser source and the measurement unit required for post-selection.

\section{DI-QKD}
As already discussed, interference-based QKD allows to establish the security of QKD as long as the functioning of the QKD transmitter is characterized, eliminating all security loopholes from the measurement unit. Note, however, that exhaustively characterizing a QKD transmitter is a difficult task (see \textit{e.g.}~\cite{pereira2019quantum,pereira2020quantum} for recent analyses). What is more, the source models are often dependent on the details of the implementation, which might also be undesirable for the purpose of standardizing QKD in the future. 

\begin{figure}[!htbp]
    \centering
    \includegraphics[width=0.76\columnwidth]{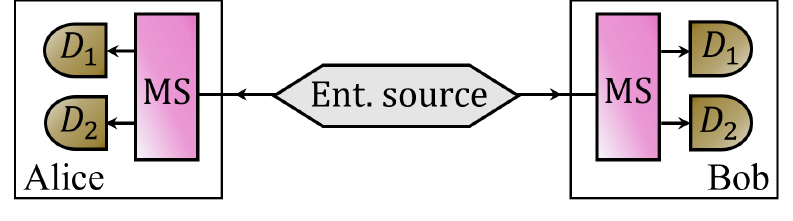}
    \caption{Schematic illustration of a DI-QKD setup. MS: measurement setting; $\{D_{j}\}$: photo-detectors. With this architecture, one can establish QKD security without internally characterizing either the entanglement source or the measurement units.\label{fig:DI}}
\end{figure}

Some of these problems can be avoided in the fully DI approach to QKD~\cite{mayers1998quantum,acin2007device,barrett2005no,primaatmaja2023security,zapatero2023advances}, a variant of entanglement-based QKD where the honest parties hold QKD receivers and an external untrusted source distributes entangled states between them (see Fig.~\ref{fig:DI}). Crucially, the DI approach completely removes the need for modelling the quantum states prepared and the measurements performed. It requires however that the QKD receivers do not reveal any information about the measurement outcomes to the outside world, which could in fact be hard to guarantee in practice given the difficulty of characterizing IL. Nevertheless, an attempt to do so has been presented in~\cite{tan2023robustness}.

Fundamentally, DI security requires that Alice's and Bob's measurement outcomes largely violate a Bell inequality~\cite{brunner2014bell}, which certifies the presence of monogamous correlations between these outcomes on the sole basis of the input-and-output measurement statistics (see \textit{e.g.}~\cite{primaatmaja2023security} for an updated review on the security of DI-QKD protocols). Importantly, the conclusiveness of a Bell test is subject to the closure of a detection loophole and a locality loophole among others (see for instance~\cite{pironio2009device}), setting stringent constraints on the tolerable loss and the timing/synchronization of Alice's and Bob's systems. Notwithstanding, various loophole-free Bell tests have been conducted in the last decade ---both photonic~\cite{giustina2015significant,shalm2015strong,li2018test} and non-photonic~\cite{hensen2015loophole,rosenfeld2017event}--- and the first proof-of-concept DI-QKD experiments have recently been reported using different platforms~\cite{nadlinger2022experimental,zhang2022device,liu2022toward}.

Remarkably though, in spite of the security upgrade, the key generation rate and the transmission distances achievable by DI-QKD in the foreseeable future are significantly limited if compared with non-DI QKD solutions~\cite{zapatero2019long,zapatero2023advances}.

\section{Malicious equipment}
While different QKD variants require different levels of physical characterization, a transversal assumption common to all of them is the \textit{honesty} of the QKD devices and the classical post-processing units, as failure to fulfil this assumption may obviously jeopardize the implementation security of QKD as well. After all, once the quantum communication of a QKD protocol ends, the resulting key data is a purely classical object subject to being copied, and a malicious manufacturer could have plenty of time and opportunities to sabotage the QKD equipment and benefit from this vulnerability. Indeed, private information might be revealed in very convoluted ways by a malicious device (be it a quantum module or a classical processor with access to the key data), and practical solutions to tackle this problem have in fact been studied~\cite{barrett2013memory,curty2019foiling,zapatero2021secure}. As an example, the techniques presented in~\cite{curty2019foiling,zapatero2021secure} are based on using redundant devices and secret sharing techniques, and a related experiment has proven their feasibility~\cite{li2021experimental}.

\section{Outlook}
The information-theoretic security of QKD is entirely rooted in the physical layer: it stems from the quantumness of the information carriers and the measurements performed on them. As a consequence, the long-term security guarantee offered by QKD comes at the price of accurately modelling the QKD equipment, and failure to do so may open practical security loopholes.

Focusing on discrete-variable QKD, in this work we have discussed many of the implementation vulnerabilities that threaten these systems, together with the suggested solutions and countermeasures. In this regard, the progress in bridging the gap between security proofs and experiments has been leaded by the refinement of both the theory and the practice of QKD, the former being the main focus of this review. Decoy-state QKD, loss-tolerant QKD, interference-based QKD or device-independent QKD are prominent examples pushing the implementation security boundaries, each of them with its own merits and limitations.

At the present time, the implementation security of QKD is one of the main concerns of governments and institutions regarding the standardization of this technology~\cite{NSA,UK_gov,FR_gov}. Apparently, the proliferation of quantum hacking attacks has undermined the idealized picture of QKD originally drawn in its early years. Notwithstanding, quantum hacking is actually helping us improve the implementation security of QKD. In fact, the combination of interference-based QKD with sufficiently isolated transmitters seems to provide an already practical solution today. On top of it, for most hardware vulnerabilities of a QKD system, either they are exploited during the key distribution session or they can never compromise the security of the delivered keys.


One way or the other, the progress in securing QKD implementations is undeniable. The gap between theory and practice has been reduced significantly and multiple applications are popping up for this disruptive technology.

\section{Acknowledgements}
This work was supported by Cisco Systems Inc., the Galician Regional Government (consolidation of Research Units: AtlantTIC), the Spanish Ministry of Economy and Competitiveness (MINECO), the Fondo Europeo de Desarrollo Regional (FEDER) through the grant No. PID2020-118178RB-C21, MICIN with funding from the European Union NextGenerationEU (PRTR-C17.I1) and the Galician Regional Government with own funding through the “Planes Complementarios de I+D+I con las Comunidades Autónomas” in Quantum Communication, the European Union’s Horizon Europe Framework Programme under the Marie Sklodowska-Curie Grant No. 101072637 (Project QSI) and the project “Quantum Security Networks Partnership” (QSNP, grant agreement No. 101114043).
\section{References}
\bibliography{References}

\end{document}